\documentclass[12pt]{article}

\usepackage{scicite}

\usepackage{times}

\usepackage{graphicx}

\newenvironment{sciabstract}{%
\begin{quote} \bf}
{\end{quote}}

\newcounter{lastnote}

\title{Beyond Stefan-Boltzmann Law: Thermal Hyper-Conductivity }

\author
{Evgenii E. Narimanov$^{1\ast}$ and Igor I. Smolyaninov$^{2}$\\
\\
\normalsize{$^{1}$Birck Nanotechnology Centre and School of Electrical and Computer Engineering}\\
\normalsize{Purdue University, West Lafayette, IN 47907, USA}\\
\normalsize{$^{2}$Department of Electrical and Computer Engineering}\\
\normalsize{University of Maryland, College Park, MD 20742, USA}\\
\\
\normalsize{$^\ast$To whom correspondence should be addressed; E-mail:  evgenii@purdue.edu.}
}

\date{}

\begin{document} 


\baselineskip24pt


\maketitle


\begin{sciabstract}
 We demonstrate that the  broadband divergence of the photonic density of states in hyperbolic metamaterials leads to giant increase in radiative heat transfer, beyond the limit set by the Stefan-Boltzmann law. The resulting  radiative thermal ``hyper-conductivity'' may approach or even exceed heat conductivity via electrons and phonons in regular solids.   \end{sciabstract}


Originally introduced to demonstrate negative refraction  \cite{Smith2003} and  to overcome the diffraction limit of optical imaging, \cite{PRB2005},\cite{OE2006}   hyperbolic metamaterials demonstrate a number of novel phenomena resulting from the broadband  singular behavior of their density of photonic states, \cite{PRL2010} which range from super-resolution imaging \cite{OE2006,Science2006,Igor2006} to enhanced quantum-electrodynamic effects \cite{APB2010,OL2010} and new stealth technology \cite{CLEO2010}.  The nature of this ``super-singularity'' in hyperbolic metamaterials can be understood from a visual representation of the density of states in terms of the phase space volume enclosed by two surfaces corresponding to different values of the light frequency.\cite{Verdeyen}  For extraordinary waves in a uniaxial dielectric metamaterial, the dispersion law
\begin{eqnarray}
\frac{k_\parallel^2}{\epsilon_\perp} + \frac{k_\perp^2}{\epsilon_\parallel} & = & \frac{\omega^2}{c^2}
\label{eq:dispersion}
\end{eqnarray}
describes an ellipsoid in the wave momentum ($k$-) space (which reduces to a sphere in isotropic media where
$\epsilon_\parallel = \epsilon_\perp$). The phase space volume enclosed between two such surfaces is then finite, corresponding to a finite density of photonic states. However, when one of the components of the dielectric permittivity tensor is negative, Eqn. (\ref{eq:dispersion}) describes a hyperboloid in the phase space. As a result, the phase space volume between two such hyperboloids (corresponding to different values of frequency) is infinite , leading to an infinite density of photonic states.
While there are many mechanisms leading to a singularity in the density of photonic states, this one is unique as (in the effective medium limit)  it leads to the infinite value of the density of states for every frequency where different components of the dielectric permittivity have opposite signs. It is this behavior which lies in the heart of the robust performance of hyperbolic metamaterials: while disorder can change the magnitude of the dielectric permittivity components, leading to a ``deformation'' of the corresponding hyperboloid in the phase (momentum) space,  it will remain a hyperboloid and will therefore still support an infinite density of states.\cite{CLEO2010} Such effective medium description will eventually fail at the point when the wavelength of the propagating mode becomes comparable to the size of the hyperbolic metamaterial unit cell $a$, introducing a natural wavenumber cut-off
\begin{eqnarray}
k_{\rm max} \sim 1/a
\label{eq:cutoff}
\end{eqnarray}
Depending on the metamaterial design and the fabrication method used, the unit cell size in optical metamaterials runs from $a \sim 10$ nm (semiconductor \cite{NatureMaterials2007} and metal-dielectric layered structures \cite{APB2010}) to $a \sim 100$ nm (nanowire composites \cite{APL2009},\cite{Science2008}). As the ``hyperbolic'' enhancement factor in the density of states \cite{PRL2010} scales as 
\begin{eqnarray}
\rho\left(\omega\right) \propto \rho_0\left(\omega\right) \left(\frac{k_{\rm max}}{\omega/c}\right)^3,
\end{eqnarray}  
where $\rho_0 \sim \omega^2$  is the free-space result, even with the cut-off taken into account, the ``hyper-singularity'' leads to the optical density of states enhancement by a factor of $10^3$ -- $10^5$.
Physically, the enhanced photonic density of states in the hyperbolic metamaterials originates from the waves with high wavenumbers that are supported by the system. Such propagating modes do not have an equivalent in ``regular'' dielectrics where $k \leq \sqrt{\epsilon} \omega/c$. 
As each of these waves can be thermally excited, a hyperbolic metamaterial will therefore show a dramatic enhancement in the radiative transfer rates.

\begin{figure}[htbp] 
   \centering
   \includegraphics[width=5in]{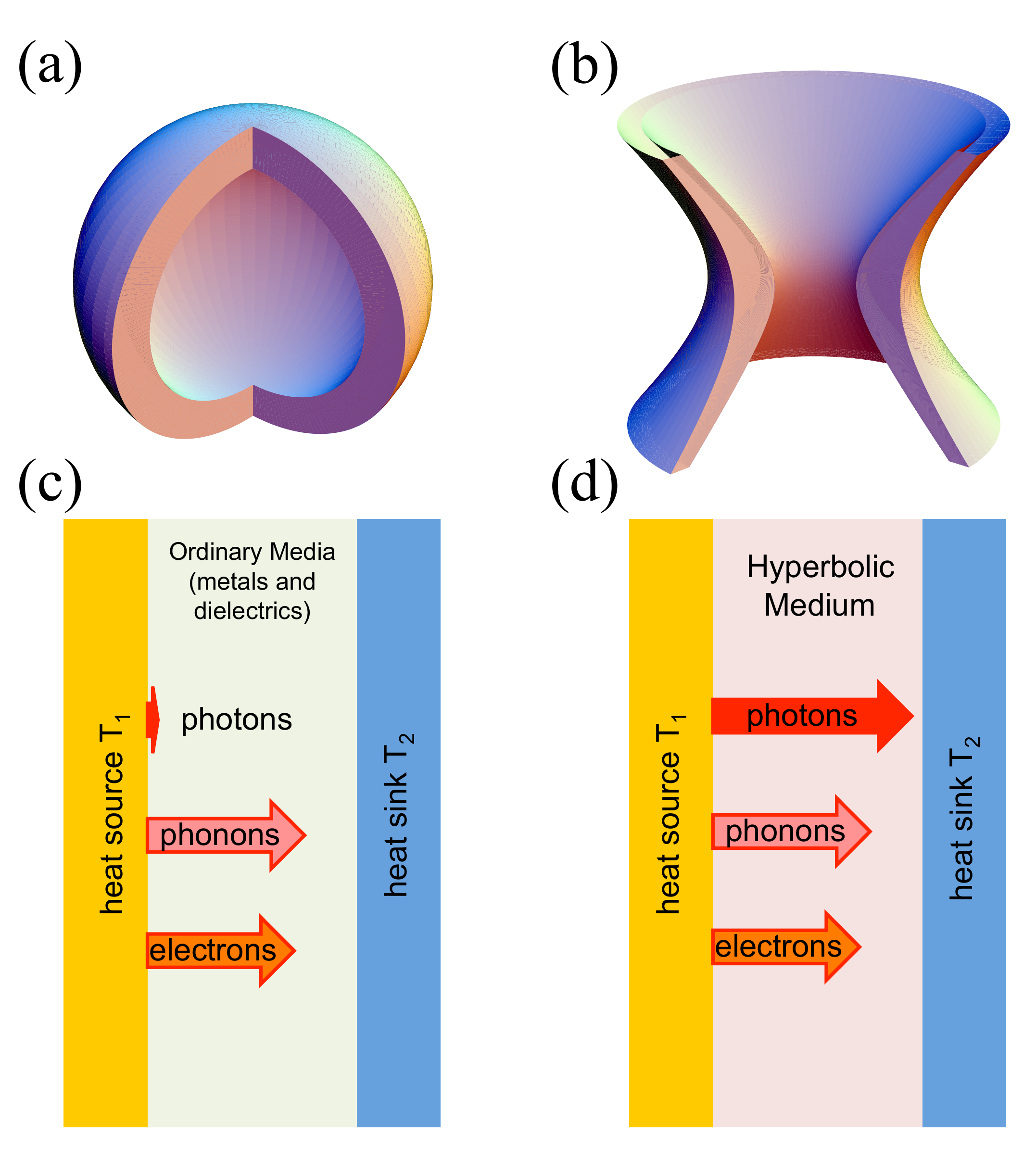} 
   \caption{The phase space volume between two constant frequency surfaces for (a) dielectric (elliptical) and (b)  hyperbolic material with $ \epsilon_\perp < 0$, $\epsilon_\parallel > 0$ (cut-out view). Panels (c) and (d) schematically  illustrate different thermal conductivity mechanisms in (c) regular media (metals and dielectric) and (d)  hyperbolic media. Giant radiative contribution to thermal conductivity in hyperbolic media can dominate the thermal transport.
}
   \label{fig:1}
\end{figure}

Furthermore, it is the density of the photonic states $\rho\left(\omega\right)$ that limits the blackbody radiation energy density $u_T$ and the energy radiated per unit area of a black body $S_T \propto u_T$
\begin{eqnarray}
u_T & = & \int_0^\infty d\omega \ \frac{\hbar \omega}{\exp\left(\frac{\hbar\omega}{kT} \right) - 1} \rho\left(\omega\right),
\end{eqnarray}
leading to the Stefan-Boltzmann upper bound to the radiative energy flux $S_T^{(0)} = n^2 \sigma T^4$ and the corresponding value  of the electromagnetic energy density $u_T^{(0)} = ( 4 n^3 / c ) \sigma T^4$ for a dielectric  with the refractive index $n$.
As a result, the singular behavior of the photonic density of states in hyperbolic metamaterial takes these media beyond the realm of the Stefan-Boltzmann law, with no ultimate limit on the radiative heat transfer.

For the energy flux along the symmetry axis of a uniaxial hyperbolic metamaterial, we find (see {\it Supplementary Materials})
\begin{eqnarray}
S_T & \simeq  &\frac{\hbar c^2 k_{\rm max}^4}{32 \pi^2} \int_{\epsilon_\parallel \cdot \epsilon_\perp <0} d\omega  \ \frac{1}{\exp\left(\frac{\hbar\omega}{k_B T}\right) - 1}\left| \   \frac{\epsilon_\perp \frac{d\epsilon_\parallel}{d\omega} - \epsilon_\parallel \frac{d\epsilon_\perp}{d\omega} }{ {\rm det} \left|\left| \epsilon \right|\right|}  \ \right|
\label{eq:S_T}
\end{eqnarray}
where the frequency integration is taken over the frequency bandwidth corresponding to the hyperbolic dispersion. 
Note that the heat flux of Eqn. (\ref{eq:S_T}) is very sensitive to the dispersion in the hyperbolic metamaterial, $d\epsilon/d\omega$. Indeed, the derivative of the dielectric permittivity determines the {\it difference} in the asymptotic behavior at $k \to \infty$ of the two hyperbolic surfaces that determine the phase space volume between the frequencies $\omega$ and $\omega + d\omega$ (see Fig. \ref{fig:1}), and thus defines the actual value of the density of states.

While there are many metamaterial designs leading to the hyperbolic dispersion, the most practical and widely used systems rely on either the medal-dielectric \cite{APB2010}  or doped-undoped semiconductor \cite{NatureMaterials2007} layer approach, or incorporate aligned metal nanowire composites \cite{APL2009},\cite{Science2008}. For the planar layers design, the hyperbolic behavior is observed for the wavelengths above 
$ \sim 10 \ \mu$m if the system is fabricated using semiconductors \cite{NatureMaterials2007},  or for the wavelength above $\sim 1 \ \mu$m if the metamaterial is composed of metal-dielectric layers \cite{APB2010}. For the nanowire-based approach, the hyperbolic dispersion is present at $\lambda \geq 1 \mu$m  \cite{APL2009},\cite{Science2008}. As a result, with either of these conventional metamaterial designs, the desired hyperbolic behavior covers the full range of wavelength relevant for the radiative heat transfer. 
We then obtain (see {\it Supplementary Materials}) 
$$S_T   \simeq   \frac{\epsilon^{(0)}}{4} S_T^{(0)}  \left(\frac{k_{\rm max}}{k_p}\right)^4$$ for the layered metamaterial design, and $$S_T    \simeq S_T^{(0)} \frac{5}{16 \pi^2}  \left(\frac{k_{\rm max}^2}{k_T k_p}\right)^2$$
 for the nanowire-based composites. Here, 
 $S_T^{(0)}$ is the blackbody thermal  energy flux for emission into the free space,
$\epsilon^{(0)} \simeq \frac{\epsilon_{\rm d}}{1 - p}$, $p$ is the volume fraction of the conducting component of the metamaterial,  $\epsilon_{\rm d}$ is the permittivity of the dielectric component of the composite, 
$
k_p  =  \sqrt{\frac{4 \pi N }{m^*}} \frac{e}{c},
$
$N$ and $m^*$ are respectively the free charge carrier density in the metamaterial and their effective mass, and the thermal momentum
$
k_T  =  {k_B T}/{\hbar c}.
$
Parametrically, the nanowires-based approach shows a higher enhancement, as $k_T \ll k_p$ ($\lambda_T \simeq 10 \ \mu$m, $\lambda_p \simeq 1 \ \mu$m). However, with existing technology metamaterial layers can be fabricated with much smaller thickness (down to $10$ nm) than the practical values for the nanowire periodicity ($\geq 100$ nm). As a result,  in both cases we find
$
S_T \simeq  \left(10^4 \ldots 10^5 \right) S_T^{(0)},
$
thus firmly placing hyperbolic metamaterials in the realm of practical applications for radiative heat transfer and thermal management.

We further note that, in addition to the thermal energy flux $S_T$,  a similar enhancement can be also observed in the thermal conductivity of hyperbolic media.\cite{ENconductivity}

In conclusion, we have examined radiative heat transfer inside hyperbolic metamaterials. We have shown that the broadband divergence of the photonic density of states leads to giant increase in radiative heat transfer compared to the Stefan-Boltzmann law in vacuum and in  dielectric materials. Our numerical results demonstrate that this radiative thermal ``hyper-conductivity'' may approach or even exceed heat conductivity via electrons and phonons, with the additional advantage of radiative heat transfer being much faster.  This radiative thermal hyper-conductivity effect may find many applications in heat management systems, especially in microelectronics.

\bigskip

 E.N. acknowledges discussions with Z. Jacob and I. Aladinski and support from C. Motti.

\bibliography{scibib}

\bibliographystyle{Science}


\end{document}